\documentclass[preprint,showpacs,preprintnumbers,amsmath,amssymb]{revtex4}

\usepackage{dcolumn}
\usepackage{bm}
\usepackage{epsfig}

\def\e{\begin{equation}}
\def\f{\end{equation}}
\def\=#1{\overline{\overline #1}}
\def\_#1{\overline #1}
\def\-#1{{\bf #1}}
\def\o{\omega}
\def\.{\cdot}
\def\M0{\mu_0}

\def\l#1{\label{eq:#1}}
\def\r#1{(\ref{eq:#1})}
\def\am{\left(\begin{array}{c}}
\def\amm{\left(\begin{array}{cc}}
\def\a{\end{array}\right)}

\begin{document}

\title{Role of wave interaction of wires and split-ring\\
resonators for the losses in a left-handed composite}



\author{C. R. Simovski}
\affiliation{Department of Physics, State Institute of Fine Mechanics and Optics\\
Sablinskaya Street, 14, 197101, St. Petersburg, Russia}

\author{B. Sauviac}
\affiliation{Laboratory DIOM, ISTASE, University Jean Monnet\\
23, Street Paul Michelon, 40042, St. Etienne, France\\
Fax: 33-4-77485039; email: bruno.sauviac@univ-st-etienne.fr}

\begin{abstract}
\noindent In this work the analytical model explaining the losses
in the known two-phase lattice of D. Smith, S. Schultz and R.
Shelby representing a uniaxial variant of left-handed medium (LHM)
at microwaves is presented. The role of electromagnetic
interaction between split-ring resonators (SRRs) and straight
wires leading to the dramatic increase of ohmic losses in SRRs
within the band when the meta-material becomes a LHM is clarified.
 {This paper explains why in this structure, rather high
transmission losses are observed in the experimental data, whereas
these losses in a separate lattice of SRRs and in a lattice of
wires are negligible at these frequencies.}
\end{abstract}

\maketitle

\subsection*{1.~Introduction}
In his seminal work \cite{Veselago68} V. Veselago summarized an
extensive study of electromagnetic properties of media with real
and {\it negative} parameters called as left-handed materials
(LHM). Since 2000 this topic has become a subject of an abundant
 {discussion initiated by } J. Pendry \cite{lens}. In 2001,
the negative refraction (a feature of LHM that allows to
distinguish these media from conventional ones) was demonstrated
in the microwave range by the group of D. Smith \cite{Shelby2001}.
Since 2001 this result has been several times reproduced (see
\cite{Lu} and \cite{Houck}) and now is considered as reliable. The
material designed in \cite{Shelby2001} is essentially a two-phase
composite, where the two phases are respectively responsible for
electric and magnetic polarization effects. The negative real part
of effective permittivity $\epsilon_{eff}$ was created by an array
of parallel conducting wires (whose direction determines the
optical axis $x$ of the composite medium), which is known to
behave similarly to a free-electron plasma at enough low
frequencies. The negative value of the real part of permeability
$\mu_{eff}$ was provided by double split-ring resonators (SRRs),
see also \cite{Pendry1999}. This negative permeability arises due
to the resonance of the magnetic polarizability of SRRs  {within a
}very narrow sub-band which belongs to their resonant band. This
sub-band lies in the wide frequency range where ${\rm
Re}(\epsilon_{eff})<0$ (more exactly the real part of a
$xx-$component of the permittivity tensor is negative). The
meta-material becomes a LHM within this sub-band for waves
polarized along $x$ and propagating orthogonally to this axis.
These electromagnetic waves suffer the magnetic and dielectric
losses. There has been a whole literature written about these
losses since the work by N. Garcia and M. Nieto-Vesperinas
\cite{Garcia}. In this work the structure tested in
\cite{Shelby2001} is treated as opaque and an exotic explanation
of negative refraction is presented. Comparing the results
obtained in the group of N. Garcia (\cite{Garcia}, \cite{Garcia1},
etc) or the results obtained in the group of A.L. Efros (e.g. in
\cite{Pokrovsky}) with the data of the group of C. Soukoulis (e.g.
in \cite{Soukoulis}) one can see that different simulations of the
same lattice give very different results for the transmission
losses per unit thickness (from practical absence to huge values).
Experimental data do not fit with all these simulations and give
moderate values for transmittance in the LHM regime. The
disagreement of the data from \cite{Soukoulis} with the experiment
is referred in this work to the influence of the dielectric board,
but it is only a guess.  {It has not been clearly stated }which
losses dominate in this meta-material: ohmic losses or dielectric
losses. However, it is clear that the losses are rather
significant and should be studied once more.

Note, that few alternative versions of a LHM at microwaves were
suggested in the literature, e.g. \cite{sim1} and \cite{book}. In
\cite{sim1} the self-consistent analytical theory of the
quasi-isotropic lattice of metal bianisotropic (Omega) particles
has been presented, and the negative parameters predicted within
the band $8.2\dots 8.4$ GHz. However, the losses were neglected in
this work. In \cite{book} the experimental testing of a LHM made
from SRRs combined with capacitively loaded strips (instead of
long wires) has been done. The aim of this work (as well as
\cite{sim1}) was to obtain an isotropic variant of LHM and to
match this medium to free space.  {The losses in this structure
are very high (the mini-pass-band corresponding to negative
material parameters is almost invisible within the band of the
resonant absorbtion of SRRs), and the band in which both material
parameters extracted from measured data have negative real parts
is rather narrow. A $50$ MHz band has been detected at $9.5$ GHz
and another one were located around $11$ GHz (however at these
frequencies the lattice period becomes longer than $\lambda/4$ and
the local constitutive parameters are not physically sound.)} In
\cite{proceedings} the transmittance through the layer of a
racemic medium from resonant chiral particles is calculated. This
medium also exhibits negative real part of constitutive parameters
within the resonant band. However, the result for resonant
transmission losses in \cite{proceedings} is pessimistic.

Therefore in the present paper we return to the structure
suggested by the group of D. Smith. We use a self-consistent
analytical model for its material parameters taking into account
ohmic and dielectric losses. The similar structure in its lossless
variant has been already studied in \cite{sim2} where the
analytical model was presented for its material parameters. The
difference of the structure suggested in \cite{sim2} with that
from \cite{Shelby2001} was another geometry of a SRR. We
considered in \cite{sim2} the SRRs suggested by R. Marques instead
of SRRs of J. Pendry. The resonator of R. Marques is not
bianisotropic, whereas the lattice of D. Smith, S. Schultz and R.
Shelby is in fact a weakly bianisotropic medium within the
resonant band of SRRs (see \cite{Marques}). The aim of \cite{sim2}
was to find the band-gap structure of the meta-material. In the
present paper we consider the SRRs of J. Pendry assuming that
these are prepared from a wire with round cross section. In
\cite{Shelby2001} the SRRs were prepared from a thin metal strip.
However, this is not a principal difference in what concerns the
scattering properties of a SRR as was clearly shown in
\cite{Bruno} and \cite{Bruno1}. Our choice of the usual wire
instead of a strip wire is explained by absence of an analytical
model of losses for curved strip wires. We consider at the first
step the lattice of parallel SRRs and at the second step we study
the meta-material from \cite{Shelby2001}. Comparing the result for
the effective permeability of two meta-materials: SRRs with wires
and SRRs only we can see the role of the electromagnetic
interaction in the two-phase material.

\subsection*{2.~Calculations of magnetic losses in a lattice of SRRs}

In the model of  the dense lattice of SRRs presented in
\cite{Pendry1999}, the calculation of ohmic losses  {did not take
into account} the curvature of wires. In this section we calculate
 {these losses}, using the Landau formula for a wire ring
\cite{Landau}. To find the permeability, we apply the rigorous
model of the dipole lattice developed in \cite{JOSA} for electric
dipoles and in \cite{PRE} for non-reciprocal magnetic dipoles. The
result we obtain in this section confirms the result from
\cite{Pendry1999}: the resonant absorption in this lattice is
rather small in the frequency band where ${\rm Re}(\mu_{eff})<0$.

We study the case of coplanar SRRs suggested in \cite{Pendry1999}.
The model of a single SRR is needed to calculate a magnetic
polarizability which enters into dispersion equation of a lattice
of magnetic dipoles derived in \cite{PRE}. The magnetic
polarizbility is defined as a relation of its magnetic moment $m$
to the local magnetic field (polarized along the $y$ axis, i.e.
orthogonally to the SRR plane)
$$
a_{mm}={m\over H^{\rm loc}}.
$$
We consider SRRs consisting of two wires with round cross section.
The model of such SRRs was developed and validated by numerical
simulations in our works \cite{Bruno,Bruno1} for the lossless
case. The conductivity resistance of the ring can be calculated
using the formula \cite{Landau} \e R_c={\rm Re}\left({2\pi r\over
\sigma'\pi r_0^2}\right). \l{rc}\f Here $r_0$ is the radius of the
wire cross section and $\sigma'$ is the effective complex
conductivity which is expressed through the metal conductivity
$\sigma$ as follows:
$$
\sigma'={2\sigma J_1(\kappa r_0)\over \kappa r_0J_0(\kappa
r_0)},\quad \kappa={(1-j)\over \delta},
$$
where $\delta=1/\sqrt{\sigma\o\mu_0}$ is the skin depth. The model
of a lossless SRR from \cite{Bruno1} is rather accurate but
sophisticated. It takes into account the non-uniformity of the
current induced in both rings and allows to calculate not only the
magnetic polarizability of a single SRR but also electric and
magneto-electric polarizabilties. Since formula \r{rc} implies the
uniform distribution of the current around the rings of SRR (and
we want to obtain a self-consistent model of losses), we put
$I^{(1,2)}=0$ in formula (47) of \cite{Bruno1}. These coefficients
describe the non-uniformity of the induced current around the
rings of SRR. When we neglect them, the long formula (12) from
\cite{Bruno1} for magnetic polarizbility of a single SRR
simplifies to a classical two-time derivative Lorentz relation
(formula (19) from \cite{Bruno1}):
 \e a_{mm}= -{ \o ^2
\M0 ^2 S^2\over L +M}{1 \over (\o^2-\o_0^2)+j\omega\Gamma }.
\l{gamma}\f $S=\pi(r_1^2+r_2^2)/2$, is the averaged area of SRR,
where $r_1$ and $r_2$ are radii of the outer and inner rings
(which are assumed to be close to one another), $L=(L_1+L_2)/2$ is
the averaged proper inductance of rings and $M$ is their mutual
inductance.

Let us first consider the lossless case. Then the factor $\Gamma$
in \r{gamma} determines radiation losses in the random medium of
SRRs \cite{Bruno} and is proportional to the radiation resistance
of the SRR, $\Gamma=\eta (k^2S)^2/6\pi(L+M)=R_{\rm rad}/(L+M)$.
Here $\eta$ is the wave impedance and
$k=\o\sqrt{\epsilon_0\epsilon_m\mu_0}$ is the wavenumber of the
host medium. Consequently, $\Gamma\sim \o^4$ (see formulae (17,20)
from \cite{Bruno1}). This result allows to our model to satisfy
the basic condition for any magnetic dipole (for electric dipoles
this condition introduced by Sipe and Kranendonk in \cite{Sipe}):
\e {\rm Im}\left\{{1\over a_{mm}}\right\}= { k^3 \over 6\pi \M0}.
\l{kappa}\f Notice, that this condition (which is obvious for both
classical and quantum scatterers and fits with the well-known
Landau correction to the Lorentz theory of dispersion) was
violated in the approximate model introduced in \cite{Pendry1999}.

Let us now study the cubic lattice of parallel lossless SRRs with
period $d$. To find $\mu_{eff}$ of the lattice we use the relation
$\mu_{eff}=\sqrt{n}$, where $n$ is the refraction index related
with the propagation factor $\beta$ of the basic propagating mode
$n=\beta/k=\beta/\o \sqrt{\epsilon_0\mu_0\epsilon_m}$. Factor
$\beta$ (within the basic Brillouin zone) can be found from the
known dispersion equation for modes propagating along $x$ or $z$
axes in a lattice of magnetic dipoles \cite{PRE}. Equation (21)
from \cite{PRE} for a simple cubic lattice of parallel magnetic
dipoles can be written in our notations as
$$ {\o\over 2\eta
d^2}{\sin kd\over \cos kd-\cos \beta d}= {\rm Re} \left(1\over
a_{mm}\right)-$$\e {\o\over 4\eta d^2}\left({\cos kR\over kR}-\sin
kR\right), \l{disp}\f where $\eta=\sqrt{\mu_0\over
\epsilon_0\epsilon_m}$ and $R\approx d/1.438$. The same equation
can be obtained from \cite{JOSA} using the duality principle. This
equation was obtained using the method of  {the} local  {field. It
gives} the band-gap structure of the lattice with real $\beta$ in
pass-bands and negative imaginary $\beta$ in stop-bands. Within
the resonant band of a scatterer when $|a_{mm}|\gg 2\eta d^2/\o$
the well-known complex mode (inherent to metallic photonic
crystals) appears and $\beta =\pi/d+j{\rm Im}\beta$. One can see
from equation \r{disp} that the radiation resistance of the
scatterer does not influence {the final parameters} of a lattice.
This results from the electromagnetic interaction in regular
structures. The imaginary part of the interaction constant of
arbitrary regular lattice exactly compensates the contribution of
the radiation resistance into inverse polarizability of scatterers
\cite{JOSA}. In \r{disp} the term ${\rm Im }(1/a_{mm})$ {has
disappeared since it } is totally compensated by the imaginary
part of the lattice interaction factor \cite{PRE}.

Now consider the case when $R_c$ is non-zero. In this case the
relation \r{kappa} naturally generalizes to \e {\rm
Im}\left\{{1\over a_{mm}}\right\}= { R_{\rm rad}+R_c \over
\o\M0^2S^2}={ k^3 \over 6\pi \M0}+{R_c \over \o\M0^2S^2}.
\l{new}\f First term of \r{new} is still totally compensated by
the interaction constant of the lattice, but the second term
remains and modifies \r{disp}. In this lossy case we should
substitute into \r{disp} the value $1/a'_{mm}$ instead of ${\rm
Re} (1/a_{mm})$. Here $a'_{mm}$ is given by \r{gamma} with the
substitution $\Gamma=R_c/(L+M)$.

\begin{figure}
\centering \epsfig{file=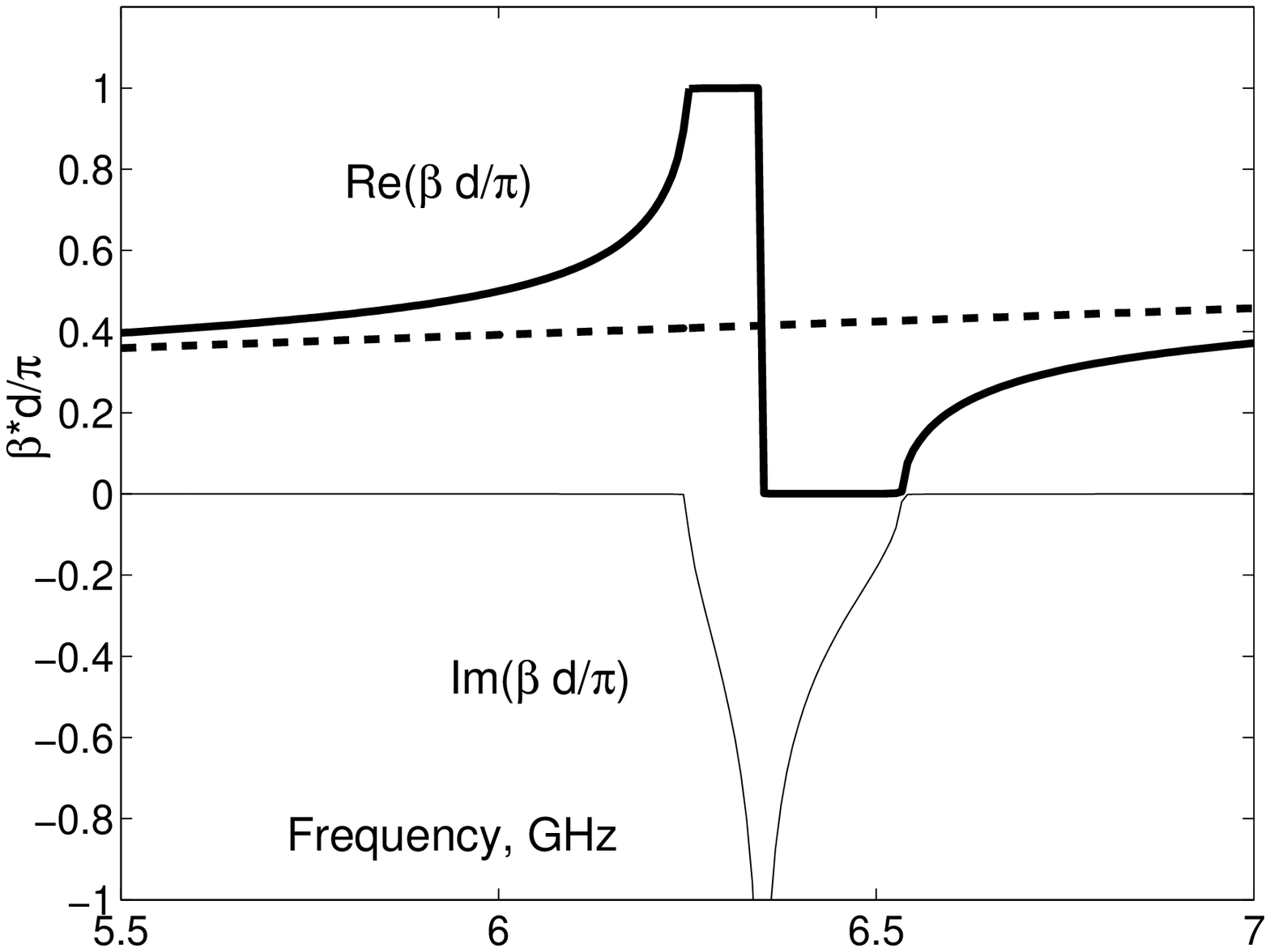, width=7cm}
\epsfig{file=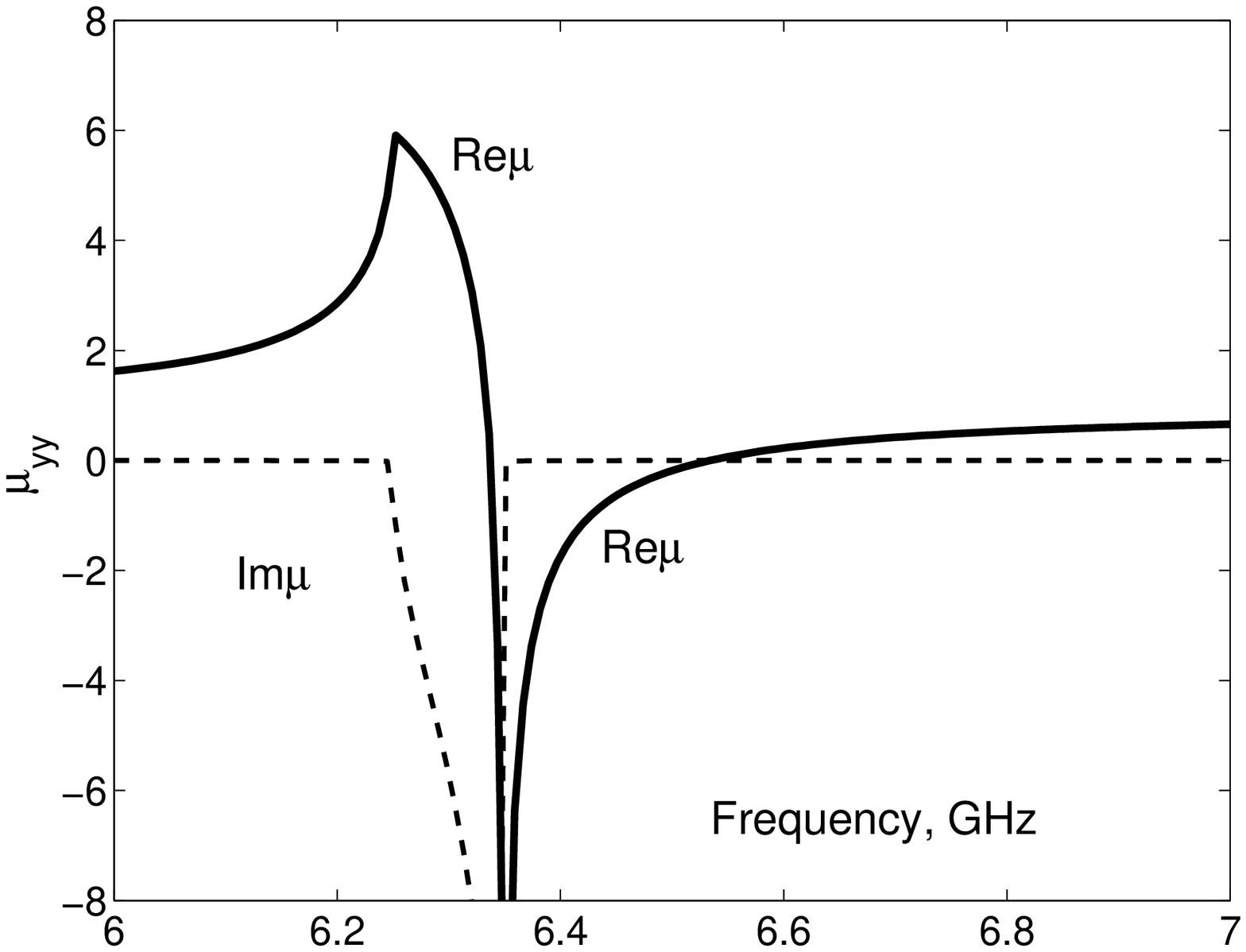, width=7cm} \caption {Top: Dispersion plot of
a cubic lattice of SRRs of copper. Bottom: Effective permeability
of lattice.} \label{fig1}
\end{figure}

Our numerical example corresponds to the following parameters:
relative permittivity of the host matrix $\epsilon_m=1.5-j0.002$,
$r_1=1.5$ mm, $r_2=1.2$ mm, wire radius $r_0=50$ mcm (it is still
$80$ times as larger as $\delta$ at $6$ GHz). Copper conductivity
$\sigma=5.8\cdot 10^7 \Omega^{-1} \cdot m^{-1} $. The period of a
lattice of SRRs was taken equal $d=8$ mm. The dispersion plot (see
Fig. \ref{fig1}, on top), shows the normalized propagation factor
$\beta d/\pi$ versus frequency and contains two  {results}. The
first one (straight dashed line) corresponds to the polarization
of electric field along $y$ and of magnetic field in the plane
$(x-z)$. Then the SRRs are not excited. The second one corresponds
to another polarization of the mode, when the SRRs are excited.
There is practically the lossless complex mode within the lower
half of the resonant frequency band of SRRs and the usual
stop-band within its upper half. The difference with the lossless
case exists but is not visible in the plot. In fact, the losses
make $\beta d/\pi$ be complex at all frequencies, however this
correction is maximally of the order $10^{-4}$.

Therefore the result for $\mu_{eff}$ presented in Fig. \ref{fig1},
on bottom, is optimistic. Within the band $6.35-6.56$ GHz we
obtained ${\rm Re}(\mu_{eff})<-0.1$ whereas $|{\rm
Im}(\mu_{eff})|<3\cdot 10^{-4}$. Though we have considered SRRs
from wires with round cross section, we have chosen  {a} very
small value for $r_0$. We expect that our results correspond to a
strip wire with width of few tenth of mm. However, we will see
below that this result cannot be expanded to the lattice of SRRs
and wires. Though the quasi-static interaction between SRRs and
wires is absent in the lattice \cite{Shelby2001},  {the wave
interaction exists}. It was clearly demonstrated in \cite{sim2}
that this interaction (in the lossless case) strongly
 {influences the real part of the material} parameters. In
the next section we will find its influence to their imaginary
parts.

\subsection*{3.~Lattice of wires and SRRs}

\begin{figure}[htb]
\centering \epsfig{file=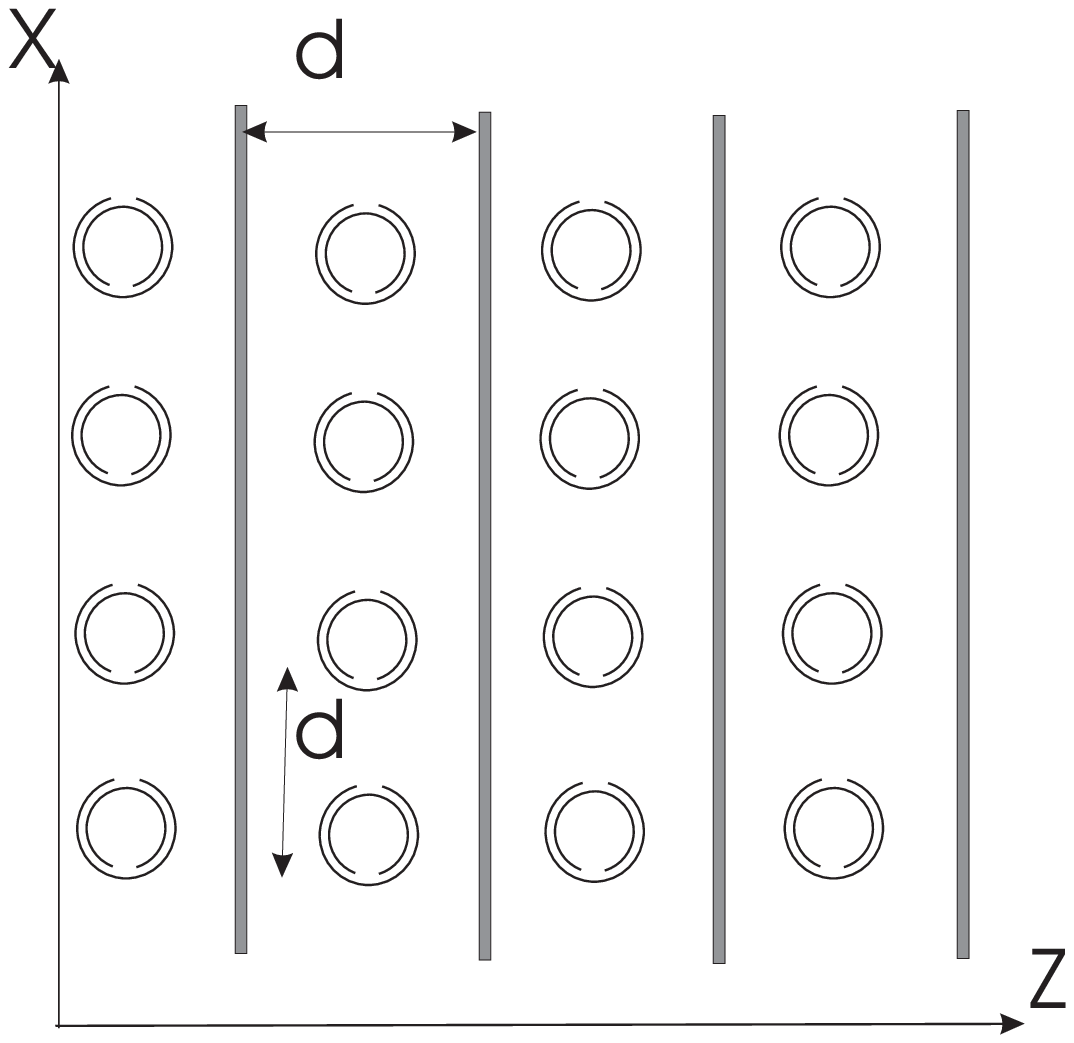, width=9cm}
\epsfig{file=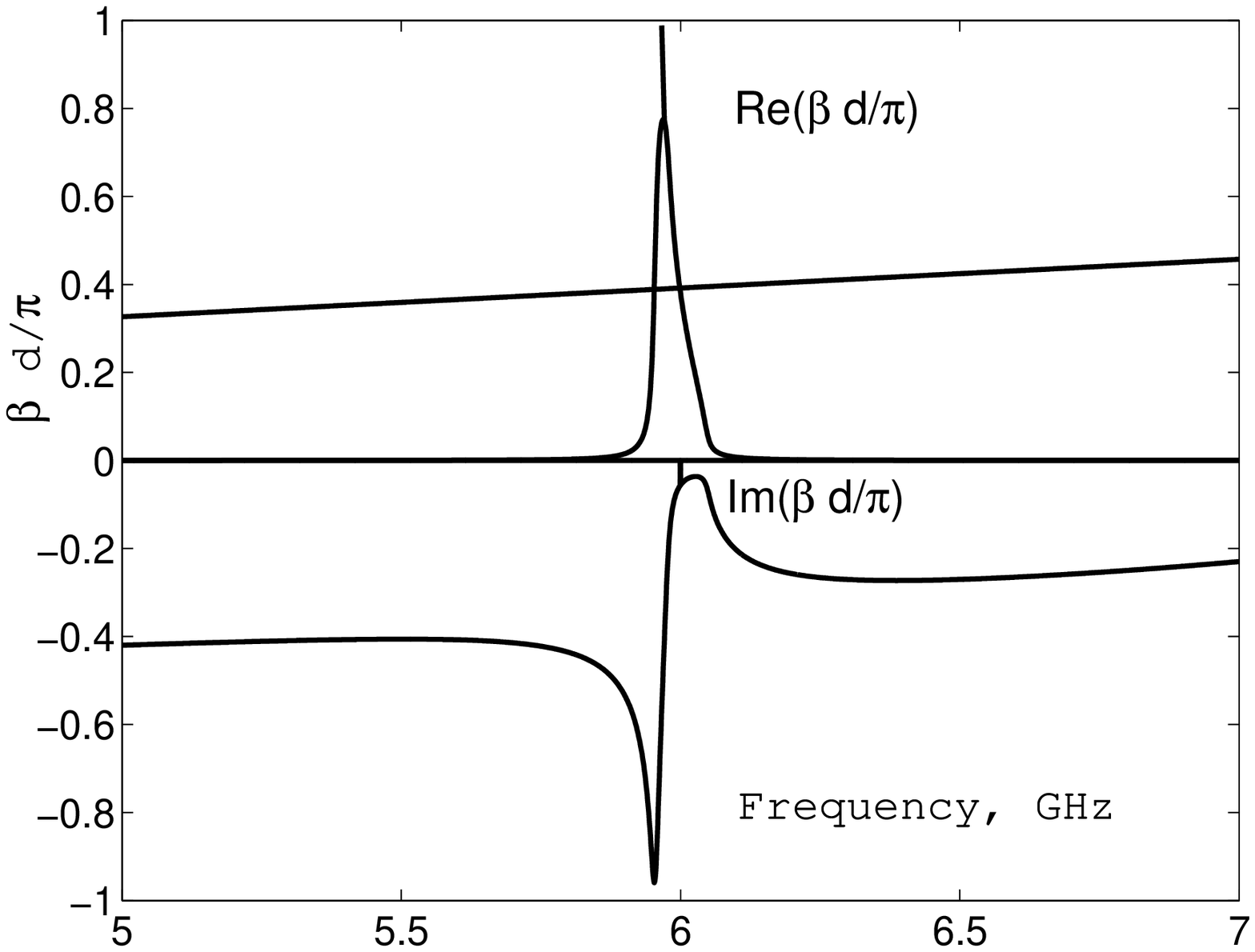, width=8cm}
 \caption {Top: Structure under study. SRRs which are parallel to the plane
 $(y-z)$ are not shown since these are not excited by the mode
under consideration propagating along $z$. Bottom: Dispersion plot
of the structure. Thin straight line corresponds to the
non-interacting mode with polarization $\-H=H\-x_0,\ \-E=E\-y_0$.}
\label{fig2}
\end{figure}

In this section we study the effective material parameters
 {of} the two-phase lattice of SRRs and wires. It is almost
the same structure as in \cite{sim2} but the metal of SRRs is not
perfect (e.g. a copper) and the background dielectric has small
losses. We assume that the wires and SRRs are located in
 {a} uniform host medium (see Fig. \ref{fig2}, on top). Our
goal is not  {to calculate the true losses in the meta-material
described in} \cite{Shelby2001} but to understand the impact of
the electromagnetic interaction  {on} these losses.  {In fact, in
\cite{sim2} the structure uses the modified SRRs of R. Marques},
whose electric polarizability is very small at the resonance of
$a_{mm}$, and magnetoelectric polarizability is zero. { In the
present paper we consider the SRRs of J. Pendry, the ones used in
\cite{Shelby2001}}. However, it does not change the dispersion
equation from \cite{sim2} since we neglect the electric and
magnetoelectric polarizabilities of SRRs.  {These are not
negligible for quantitative calculations but are not important
enough for our purposes.} All we need to modify  {in the theory}
is to add the term $R_c^t$ to the radiation resistance of a SRR
denoted as $R_r$  {in formula (10) of \cite{sim2}}. Repeating the
steps which led  {from equation (10) to (29)}, we obtain the
following dispersion equation for waves propagating along the
$z-$axis
$$ \left[{2d^2\over \o\eta}{1\over a'_{mm}}-
{1\over 2}\left({\cos kR\over kR}-\sin kR\right)\right](\cos kd-\cos
\beta d)\sin kd$$ \e -\left(1+{kd\over \pi}\log {d\over 2\pi
r_w}\right)\sin^2 kd-\cos ^2\beta d-1=0.
 \l{eq}\f
Relation \r{eq}  {results of \cite{sim2} }with the only
substitution ${\rm Re}(1/a_{mm})\rightarrow 1/a'_{mm}$. In \r{eq}
$r_w$ is the radius of straight wires. Longitudinal component of
the effective permittivity $\epsilon_{xx}$ and $\mu_{yy}$ are
determined (respectively) by formulas (35) and (36) of
\cite{sim2}.

 In our numerical example, the lattice of SRRs is the same
as in the precedent section and it is assumed that $r_w=r_0$. The
result for permeability dramatically  {differs} from the
 {previous result ; this due to the electromagnetic}
interaction of SRRs and wires. This interaction is described by
the parameter $B$ in equations (27) and (28) of \cite{sim2}. If
one formally puts $B=0$,  {the problem splits} in two independent
dispersion equations, one for the medium of SRRs and one for the
wire medium. Then $\mu_{eff}$ keeps the same values as in Fig.
\ref{fig1}.

 On Fig. \ref{fig2}, we have shown the dispersion plot of
the structure. This plot corresponds to same permittivity of host
medium $\epsilon_m=1.5-j0.002$ as in the previous section.
Comparing this plot with similar plot for lossless structure (Fig.
5 of \cite{sim2}) one finds that the complex mode (in Fig. 5 it
corresponds to the lower half of the resonant band of SRRs)
disappears. Instead, the forward wave with strong attenuation
appears in this resonant sub-band. This forward wave corresponds
to ${\rm Re}(\mu_{eff})>0$ and ${\rm Re}(\epsilon_{eff})<0$, and
the propagation is possible due to the complexity of constitutive
parameters. The attenuation factor is larger than the propagation
one in this band. The upper half of the resonant band of SRRs
contains the backward wave as well as in Fig. 5 of \cite{sim2}.
The ohmic losses broaden the frequency band of this backward wave,
however these also produce the visible imaginary part of the
propagation factor in this band. This imaginary part is not so
high as in the lower half of the resonant band and is related with
ohmic losses in SRRs. It is practically not affected by imaginary
part of $\epsilon_m$ (until the threshold $\epsilon_m=1.5-j0.1$,
when the influence of the dielectric losses in the matrix becomes
visible in the dispersion plots). In this sub-band ${\rm
Im}(\beta)$ decreases if one increases the wire radius $r_0$ and
increases if one decreases $\sigma$.

\begin{figure}[htb]
\centering \epsfig{file=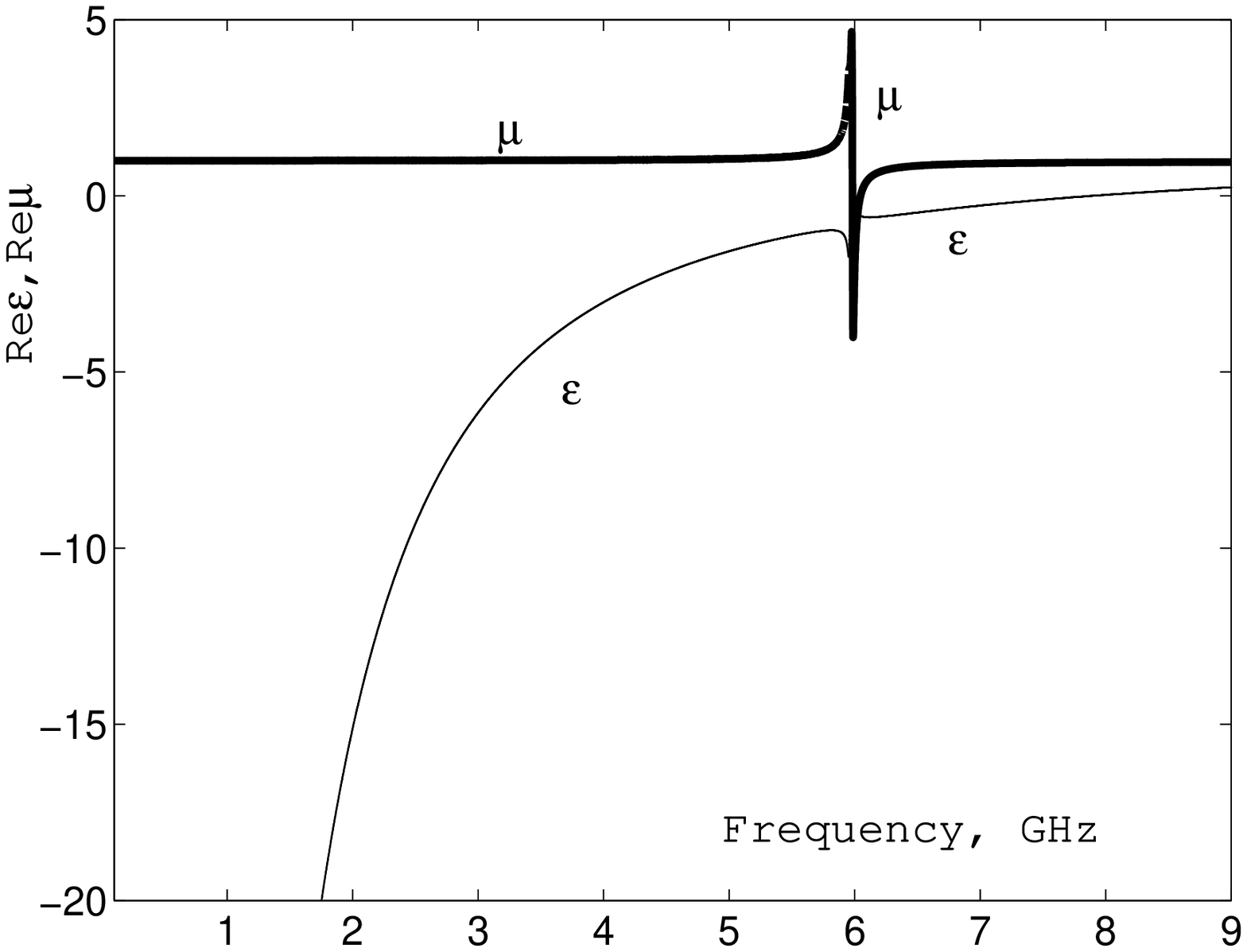, width=7.5cm}
\epsfig{file=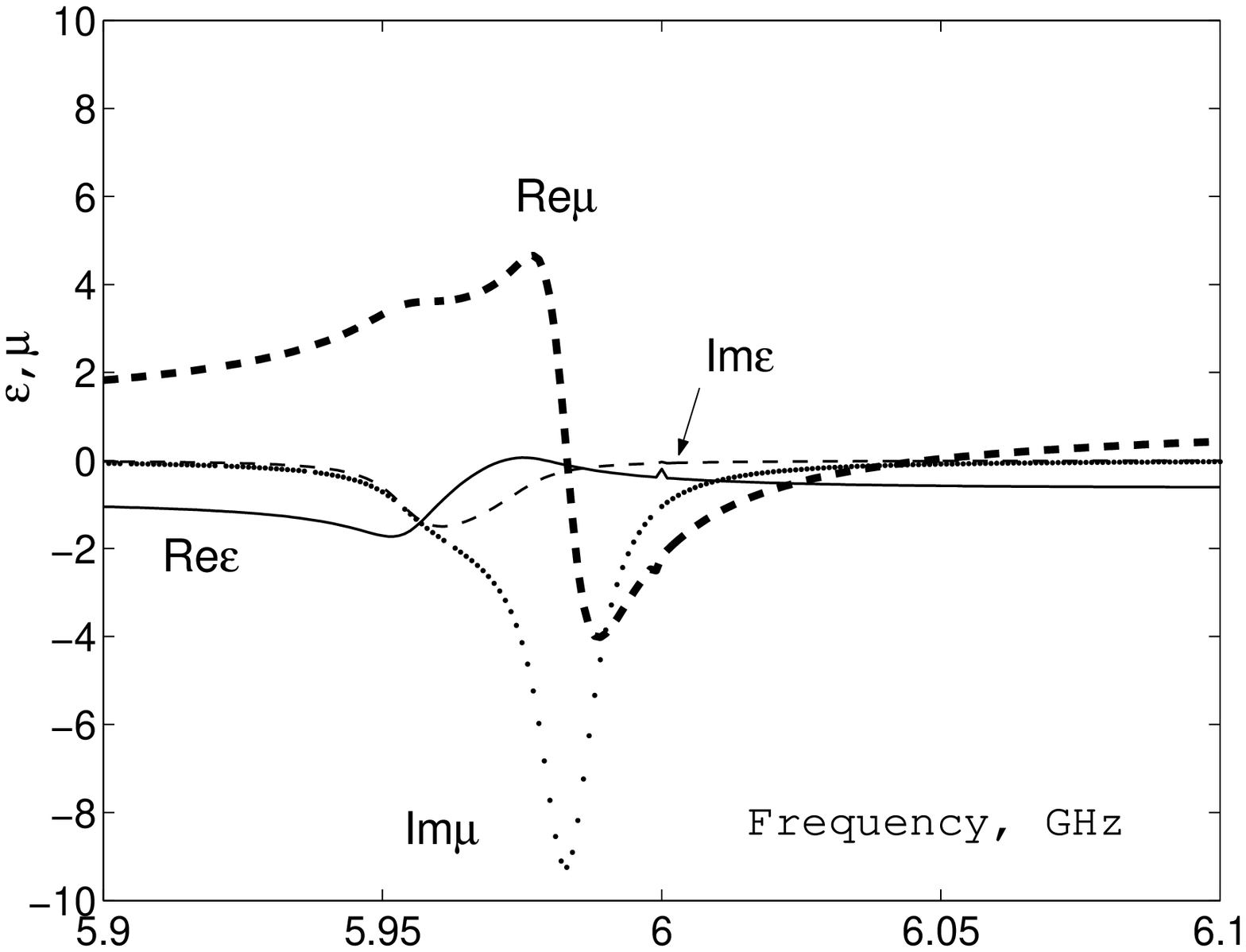, width=7.5cm} \caption {Top: Real parts
of $\epsilon_{eff},\mu_{eff}$ in the wide frequency range. Bottom:
Resonant frequency band. Real and imaginary parts of
$\epsilon_{eff}$ and $\mu_{eff}$.} \label{eq1}
\end{figure}

In Fig. \ref{eq1} we present the frequency behavior of
$\epsilon_{eff}$ and $\mu_{eff}$ for the structure under study.
Here $\epsilon_{eff}$ means the $xx-$component of tensor
$\=\epsilon$, and similarly $\mu_{eff}\equiv \mu_{yy}=\mu_{zz}$.
On top the wide-band frequency dependence is shown for real parts
of $\epsilon_{eff}$ (thin line) and $\mu_{eff}$ (thick line).
 {One can see that both parameters are resonant. This is the result of
 the electromagnetic interaction between wires and SRRs (which also
 leads to the small shift of the resonant frequency for $\mu_{eff}$
 compared to that of a single lattice of SRRs).
 Though the resonance of permittivity is weak, it remains negative within} the
resonant band of SRRs.  {This resonance confirms once more that
the very simplistic approach, in which the constitutive parameters
of the meta-material are obtained with the trivial superimposition
of $\mu$ of SRRs and $\epsilon$ of wires, is not applicable in
most of cases }\cite{sim2}.   {On the bottom of Fig. \ref{eq1},
both real }and imaginary parts of these parameters are presented
within the resonant band of SRRs. The band of the backward wave
(the upper sub-band of the resonant band of SRRs) is  {the one} in
which the meta-material becomes the LHM.  {Unlike in the
precedent} section, in this sub-band the absolute value of ${\rm
Im}(\mu_{eff})$  {is quite high}. We obtained $|{\rm
Im}(\mu_{eff})|\approx (0.1\dots 0.2) |{\rm Re}(\mu_{eff})|$
within the frequency band of the backward wave (i.e. of the
negative material parameters). This result explains the
experimental data for transmittance that can be found in the
literature.

Note, that introducing the finite conductivity for straight wires
practically does not change the result. The ohmic losses in
straight wires do not pose a difficult problem since the negative
permittivity of wire lattice is not resonant. But the permeability
of the lattice of SRRs is resonant and it makes possible the
strong influence of wires to SRRs. This is the reason of high
magnetic losses in the backward-wave regime.

\subsection*{4.~Conclusion}

In this paper we considered the problem of the influence of ohmic
losses in the metal and dielectric losses in the host matrix to
the magnetic and dielectric losses in the two-phase composite
medium formed by a square lattice of infinitely long parallel
wires and a cubic lattice of symmetrically located SRRs. It has
been shown that the electromagnetic interaction between wires and
SRRs together with ohmic losses in SRRs give rather significant
magnetic losses. These losses are absent from the single lattice
of SRRs. This way we suggest an explanation of the rather
significant transmission losses in the structure from
\cite{Shelby2001} which is considered as a uniaxial variant of
left-handed material. We do not deny the significance of the
losses in the dielectric board, which support the SRRs prepared
from copper strips. However, the contribution of ohmic losses is
also rather important.

We have shown that the wave interaction of electric and magnetic
components of the meta-material is harmful for its transmittance.
It is impossible to avoid this interaction. The transmittance can
be improved reducing the ohmic resistance of SRRs. Increasing the
width of the strip wire or the radius of the usual wire one can
obtain the better transmission properties of the meta-material
over the whole resonant band in spite of the electromagnetic
interaction in the meta-material.


\end{document}